\documentstyle[referee,psfig]{mn}
\newif\referee
\newif\ifAMStwofonts

\catcode`\@=11
\def\gsim{\ifmmode{\mathrel{\mathpalette\@versim>}}
    \else{$\mathrel{\mathpalette\@versim>}$}\fi}
\def\lsim{\ifmmode{\mathrel{\mathpalette\@versim<}}
    \else{$\mathrel{\mathpalette\@versim<}$}\fi}
\def\@versim#1#2{\lower 2.9truept \vbox{\baselineskip 0pt \lineskip
    0.5truept \ialign{$\m@th#1\hfil##\hfil$\crcr#2\crcr\sim\crcr}}}
\catcode`\@=12

\def\mlk{\Upsilon_{\rm *,K}} 
 
\def\Mmin{M_{\rm min}}
\def\Mcl{M_{\rm cl}} 
\def\Mcltot{M_{\rm cl,tot}} 
\def\Mhalo{M_{\rm h,tot}}
\def\Mstar{M_{\rm *,tot}}

\def\Mbh{M_{\rm BH}} 
\def\Ncl{N_{\rm cl}} 
\def\Nstar{N_*}
\def\Nhalo{N_{\rm h}} 
\def\Nmerg{N_{\rm merg}}

\def\Psicl{\Psi_{\rm cl}} 
\def\rhocl{\rho_{\rm cl}}
\def\rhoM{\rho_{\rm M}} 
 
\def\Lk{L_{\rm K}}

\def\Msol{M_{\odot}} 
 
\def\mue{\langle \mu \rangle _{\rm e}}
\def\Re{R_{\rm e}}
\def\cRe{\langle R\rangle _{\rm e}}
\def\ciR{\langle R \rangle}
\def\Dmu{\langle \Delta \mu \rangle}
 
\def\Dtmax{\Delta t_{\rm max}} 
\def\Dtmin{\Delta
t_{\rm min}}

\def\rbh{r_{\rm BH}} 
 
\def\rccl{r_{\rm cl}} 
\def\rchalo{r_{\rm h}} 
\def\rcstar{r_{\rm *}} 
\def\ra{r_{\rm a}}
\def\racl{r_{\rm a,cl}} 
\def\rme{r_{\rm M}} 
 
\def\sast{s_{\rm a,*}}
\def\sacl{s_{\rm a,cl}} 
\def\sg0{\sigma_0} 
\def\sgv{\sigma_{\rm V}}

\def\Td{T_{\rm dyn}} 
\def\Tfric{T_{\rm fric}} 
\def\Tlast{T_{\rm last}}

\def\en{{\mathcal{E}}}
\def\Zsun{Z_{\odot}} 
\def\psit{\Psi_{\rm T}}

\def\alphatol{\alpha_{\rm tol}}
\def\res{\mathcal{R}}
\ifoldfss
  \ifCUPmtlplainloaded \else
    \NewTextAlphabet{textbfit} {cmbxti10} {}
    \NewTextAlphabet{textbfss} {cmssbx10} {}
    \NewMathAlphabet{mathbfit} {cmbxti10} {} 
    \NewMathAlphabet{mathbfss} {cmssbx10} {} 
  \fi
  \ifAMStwofonts
    \ifCUPmtlplainloaded \else
      \NewSymbolFont{upmath} {eurm10}
      \NewSymbolFont{AMSa} {msam10}
      \NewMathSymbol{\upi}     {0}{upmath}{19}
      \NewMathSymbol{\umu}     {0}{upmath}{16}
      \NewMathSymbol{\upaellitrtial}{0}{upmath}{40}
      \NewMathSymbol{\leqslant}{3}{AMSa}{36}
      \NewMathSymbol{\geqslant}{3}{AMSa}{3E}

      \let\leq=\leqslant 
       
    \fi
  \fi
\fi 

\ifnfssone
  \newmathalphabet{\mathit}
  \addtoversion{normal}{\mathit}{cmr}{m}{it}
  \addtoversion{bold}{\mathit}{cmr}{bx}{it}
  \newmathalphabet{\mathbfit} 
  \addtoversion{normal}{\mathbfit}{cmr}{bx}{it}
  \addtoversion{bold}{\mathbfit}{cmr}{bx}{it}
  \newmathalphabet{\mathbfss} 
  \addtoversion{normal}{\mathbfss}{cmss}{bx}{n}
  \addtoversion{bold}{\mathbfss}{cmss}{bx}{n}
  \ifAMStwofonts
    \ifCUPmtlplainloaded \else
      \UseAMStwoboldmath
      \makeatletter
      \new@mathgroup\upmath@group
      \define@mathgroup\mv@normal\upmath@group{eur}{m}{n}
      \define@mathgroup\mv@bold\upmath@group{eur}{b}{n}
      \edef\UPM{\hexnumber\upmath@group}
      \new@mathgroup\amsa@group
      \define@mathgroup\mv@normal\amsa@group{msa}{m}{n}
      \define@mathgroup\mv@bold\amsa@group{msa}{m}{n}
      \edef\AMSa{\hexnumber\amsa@group}
      \makeatother
      \mathchardef\upi="0\UPM19
      \mathchardef\umu="0\UPM16
      \mathchardef\upartial="0\UPM40
      \mathchardef\leqslant="3\AMSa36
      \mathchardef\geqslant="3\AMSa3E

      \let\leq=\leqslant 

    \fi
  \fi
\fi 

\ifnfsstwo
  \DeclareMathAlphabet{\mathbfit}{OT1}{cmr}{bx}{it}
  \SetMathAlphabet\mathbfit{bold}{OT1}{cmr}{bx}{it}
  \DeclareMathAlphabet{\mathbfss}{OT1}{cmss}{bx}{n}
  \SetMathAlphabet\mathbfss{bold}{OT1}{cmss}{bx}{n}
  \ifAMStwofonts
    \ifCUPmtlplainloaded \else
      \DeclareSymbolFont{UPM}{U}{eur}{m}{n}
      \SetSymbolFont{UPM}{bold}{U}{eur}{b}{n}
      \DeclareSymbolFont{AMSa}{U}{msa}{m}{n}
      \DeclareMathSymbol{\upi}{0}{UPM}{"19}
      \DeclareMathSymbol{\umu}{0}{UPM}{"16}
      \DeclareMathSymbol{\upartial}{0}{UPM}{"40}
      \DeclareMathSymbol{\leqslant}{3}{AMSa}{"36}
      \DeclareMathSymbol{\geqslant}{3}{AMSa}{"3E}

      \let\leq=\leqslant 

    \fi
  \fi
\fi 

\ifCUPmtlplainloaded \else
  \ifAMStwofonts \else 
    \def\upi{\pi}
    \def\umu{\mu}
    \def\upartial{\partial}
  \fi
\fi

\title[Galactic cannibalism  in the cluster C0337-2522 at
$z=0.59$]{Galactic cannibalism in the galaxy cluster C0337-2522 at
${\bf z=0.59}$\thanks{This paper is partially based on data collected
at the European Southern Observatory Very Large Telescope at Paranal
(proposals 63.O-0591 and 64.O-0298)}}

\author[C. Nipoti, M. Stiavelli,  L. Ciotti, T. Treu and P. Rosati]
 {C.~Nipoti$^{1,2}$, M.~Stiavelli$^2$, L.~Ciotti$^{1,3}$, T.~Treu$^4$,
 and P.~Rosati$^5$\\ $^1$Dipartimento di Astronomia, Universit\`a di
 Bologna, via Ranzani 1, 40127 Bologna, Italy\\ $^2$Space Telescope
 Science Institute, 3700 San Martin Drive, Baltimore, MD 21218\\
 $^3$Scuola Normale Superiore, Piazza dei Cavalieri 7, 56126 Pisa,
 Italy\\ $^4$California Institute of Technology, Astronomy Department,
 MS 105-24, Pasadena, CA 91125\\ $^5$European Southern Observatory,
 Karl-Schwarzschild-Strasse 2, 85748 Garching, Germany}

\begin{document}
\vskip 1. truecm

\date{June 4, 2003 accepted}

\pubyear{2003}

\maketitle

\label{firstpage}

\begin{abstract}

According to the galactic cannibalism model, cD galaxies are formed in
the center of galaxy clusters by merging of massive galaxies and
accretion of smaller stellar systems: however, observational examples
of the initial phases of this process are lacking. We have identified
a strong candidate for this early stage of cD galaxy formation: a
group of five elliptical galaxies in the core of the X-ray cluster
C0337-2522 at redshift $z=0.59$.  With the aid of numerical
simulations, in which the galaxies are represented by N-body systems,
we study their dynamical evolution up to $z=0$; the cluster dark
matter distribution is also described as a N-body system.  We explore
the hypothesis that some of the five galaxies will have merged before
$z=0$, making reasonable assumptions on the structural and dynamical
characteristics of the cluster. We then compare the properties of the
merger remnant with those of real ellipticals (such as its accordance
with the Fundamental Plane, the Faber-Jackson, and the $\Mbh$-$\sg0$
relations) and, in particular, we check whether the remnant has the
surface brightness profile typical of cD galaxies. We find that a
multiple merging event in the considered group of galaxies will take
place before $z=0$ and that the merger remnant preserves the
Fundamental Plane and the Faber--Jackson relations, while its behavior
with respect to the $\Mbh$-$\sg0$ relation is quite sensitive to the
details of black hole merging. However, the end--products of our
simulations are more similar to a ``normal'' giant elliptical than to
a cD galaxy with its characteristic diffuse luminous halo, thus
confirming previous indications that the formation of cD galaxies is
not a necessary consequence of galaxy merging at the cluster center.

\end{abstract}


\section{Introduction}

Among the various scenarios proposed for the formation of the
brightest cluster galaxies (BCGs), and in particular of cD galaxies,
perhaps the most prominent is the so--called ``galactic cannibalism''
model (Ostriker \& Tremaine 1975, Hausman \& Ostriker 1978). In this
picture super-luminous ellipticals (hereafter Es) are formed in the
center of galaxy clusters by merging of massive galaxies and by
accretion of smaller stellar systems. Indeed, numerical simulations
have shown that galactic cannibalism is able to reproduce many
properties of the observed BCGs (see, e.g., Miller 1983, Merritt 1984,
Malumuth \& Richstone 1984, Bode et al. 1994, Athanassoula, Garijo \&
Garcia Gomez 2001). However, there are still some significant
discrepancies between the predictions of these simulations and the
observations: in particular, if one considers the available numerical
simulations it is apparent that only a small fraction of the
simulations ending in a merging event do produce a cD-like galaxy,
while observations reveal that the presence of a cD galaxy is quite a
common property of galaxy clusters (see, e.g., Dressler 1984, and
references therein). In addition, Thuan \& Romanishin (1981) pointed
out that BCGs in poor clusters do not show the diffuse luminous
envelope typical of cDs, and this could be an important additional
indication that the end--product of galactic cannibalism not
necessarily consists in a cD galaxy.

Observational tracers of the {\it late stages} of cD galaxy formation
have already been found: for example, the high frequency of multiple
nuclei in cD galaxies is considered an indication of recent merging
(see, e.g., Matthews, Morgan \& Schmidt 1964, Schneider, Gunn \&
Hoessel 1983, Laine et al. 2003). On the contrary, observational
examples of the {\it initial stages} of cD galaxy formation are
lacking: according to the scenario depicted above, such systems would
appear as groups of galaxies located in the core of clusters and
spiraling, as an effect of dynamical friction, towards the cluster
center. We have identified a strong candidate for such an evolutionary
stage: a group of five Es located within a region of a few kpc of
(projected) linear size near the center of the X-ray selected cluster
C0337-2522 at redshift $z=0.59$ (ROSAT Deep Cluster Survey; Rosati et
al. 1998).

The main goal of this work is to explore, as a function of the initial
conditions and of the structural and dynamical characteristics of
their parent cluster, how many (if any) of the five galaxies under
consideration will have merged before $z=0$. In addition, we
investigate whether the remnant has the surface brightness (SB)
profile typical of cD galaxies, i.e., it possesses a diffuse low
luminosity halo. Finally, we check whether some features of Es, such
as the Fundamental Plane (FP; Djorgovski \& Davis 1987; Dressler et
al. 1987), the Faber-Jackson relation (FJ; Faber \& Jackson 1976), the
$\Mbh$-$\sg0$ relation (Gebhardt et al.  2000; Ferrarese \& Merritt
2000) and the metallicity gradient (Peletier 1989, Carollo, Danziger
\& Buson 1993), are preserved during the process. This kind of
investigation is motivated by the fact that BCGs, both giant Es and
cDs, follow quite closely the FP and the FJ relation determined by
less luminous Es (see, e.g., Oegerle \& Hoessel 1991), though there
are indications that a significant fraction of BCGs are brighter than
would be expected from the FJ relation of normal Es (Malumuth \&
Kirshner 1981, 1985).  BCGs have also metallicity gradients consistent
with those of normal Es (Fisher, Franx \& Illingworth 1995), and those
for which the mass of the central supermassive black hole (BH) has
been measured do follow the $\Mbh$-$\sg0$ relation (e.g., M87;
Gebhardt et al.  2000, Ferrarese \& Merritt 2000).

We try to address the questions above with the aid of numerical
simulations in which the galaxies are represented by N-body systems,
and the initial conditions are constrained by the imaging and
kinematic information from our ESO-VLT data.  The observations
provide, for each galaxy, only three phase--space coordinates (the two
projected positions and the line--of--sight velocity); thus, some
assumptions are needed in order to assign the remaining initial
conditions. In principle, one could make use of non-parametric
estimators (see Merritt \& Trembley 1994); however, given the small
number of objects involved, such estimators are not practical for our
application.  To overcome this problem, we modeled the cluster where
the five galaxies reside as a spherical dark matter (DM) density
distribution with adjustable total mass, scale length and amount of
radial orbital anisotropy in the velocity distribution: in order to
explore the effects of the dynamical friction of the galaxies against
the cluster DM, also the cluster is represented as a {\it live} N-body
system.

This paper is organized as follows. The observations and the galaxy
models are presented in Section~2 and Section~3, respectively. A
description of the simulations is given in Section~4 and the results
are discussed in Section~5. Our conclusions are presented in
Section~6.

\section{Observations and data reduction}

\begin{figure}
\begin{center}
\parbox{1cm}{ \psfig{file=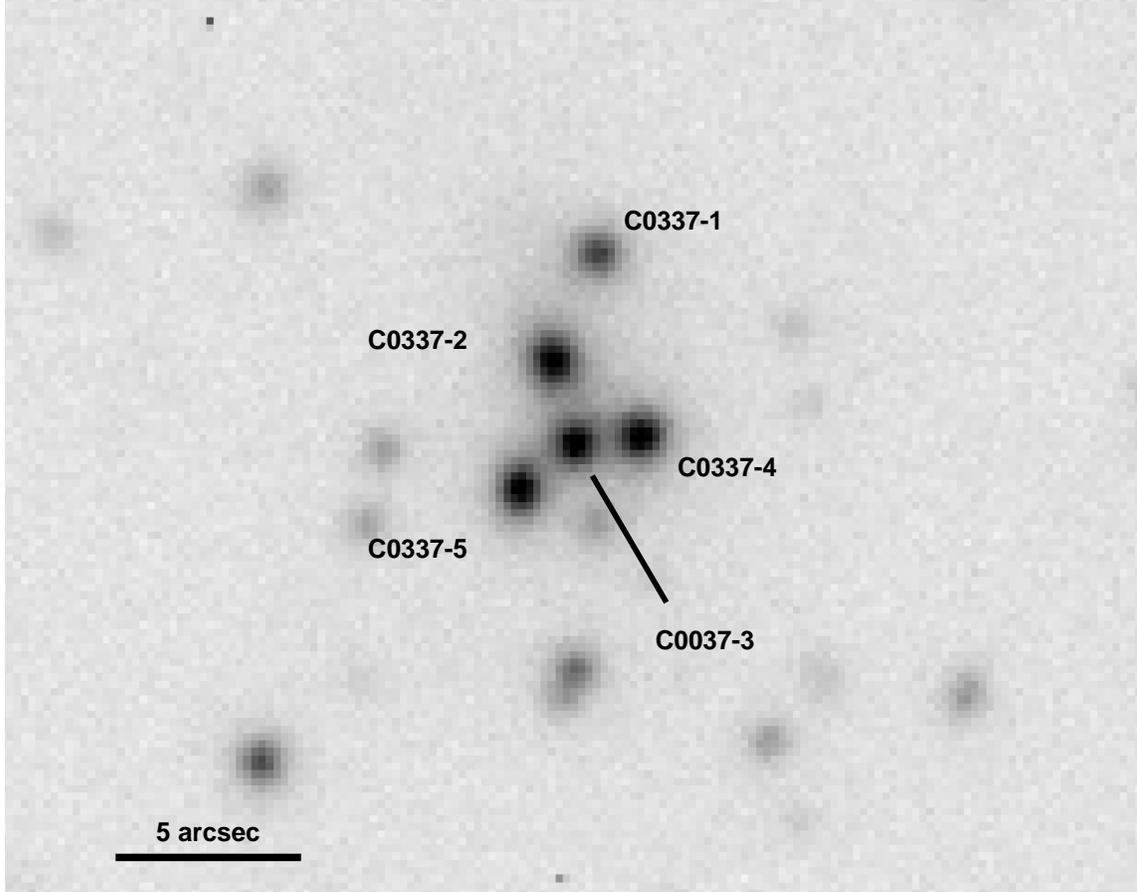,height=12cm}}
\caption{VLT-FORS1 I-band image of the galaxy cluster C0337-2522 (N up, E
left). The five galaxies are named as in Table~1.}
\label{appenfig}
\end{center}
\end{figure}

The five galaxies that we have considered are located in the core of
the X-ray galaxy cluster C0337-2522. The I-band image, shown in
Fig.~1, was obtained in September 1999, using the Focal Reducer and
Spectrograph 1 (FORS1) at the ESO Very Large Telescope (VLT) with
exposure time $2 \times 300$s and seeing $0.9''$. The image was
reduced with standard procedure and a catalogue of the objects was
carried out with the SExtractor package (Bertin \& Arnouts 1996).
Spectra for the five elliptical galaxies were obtained in January 2000
with FORS1 at VLT, using the grism R600 and a $1''$ wide slit with a
resolution of $\approx 90\, -\, 100\, {\rm km\,s^{-1}}$; exposure
times were in the range $2700\, -\, 7200\,{\rm s}$. The reduction of
the spectroscopic data and the measurement of the redshifts (reported
in Table~1) were performed following the procedures described in Treu
et al. (2001). We note that, due to poor weather conditions and
reduced reflectivity during the early-stages of operations (see, e.g.,
Labb\'e et al. 2003), the S/N ratio of the spectroscopic data does not
allow us to measure with sufficient accuracy the central velocity
dispersions for all the five galaxies. Thus, in the present work we
are unable to use velocity dispersion measurements in order to
constrain the total mass of the galaxies and their internal dynamics.
The details of the photometric measurements will be presented in a
separate paper (Treu et al. 2003). For the aim of this paper it
suffices to say that the five galaxies have very similar (within $\sim
0.5$ mag, see Table 1) I-band magnitudes, typical of bright cluster
E/S0 at that redshift (see, e.g., Kelson et al. 1997). We have no 
information about other low--luminosity members of the cluster, and so
we decided to consider only the five galaxies in the simulations,
except for a special simulation described in Section~6. Note also that
some doubt about the cluster membership of galaxy C0337-2 could be
reasonably raised by considering its negative and high barycentric
velocity: however, the cluster X-ray emission (Rosati et al. 1998; see
also Vikhlinin et al. 1998) strongly points towards the physical
association of the central group of galaxies. We decided to consider
also galaxy C0337-2 as a member of the cluster: its ``special''
dynamical status will be however evident when discussing the results
of the simulations.
\begin{table*}
 \centering
  \caption {Known data for the five galaxies.  Positions and
velocities are relative to the center of mass of the system. For
reference, the center of galaxy C0337-4 is at $\alpha=03{\rm h}37{\rm
m}45.15{\rm s}$ and $\delta=-25^o22'36\farcs1$ (J2000). I-band
aperture photometry within a $1''$ diameter aperture together with its
random uncertainty is listed in the last column (the zero point of the
magnitude scale is accurate to 0.1 mags).}
  \begin{tabular}{@{}cccccccc@{}}
    Galaxy    & redshift &  $x$ & $y$  & $v_z$ & $x$ & $y$ & I    \\
              &          & (arcsec) &  (arcsec) & (km/s) & (kpc) & (kpc) & (mag)    \\
    [10pt] 
    C0337-1  & 0.589	&  0.51 &  3.78  &   284.7 &  3.63   &   27.00  & 20.55 $\pm$ 0.05  \\
    C0337-2  & 0.578    & -0.36 &  0.94  & -1892.9 &  -4.51  &    6.71  & 20.02 $\pm$ 0.06  \\
    C0337-3  & 0.590    & -0.01 & -1.24  &   473.3 &  -0.09  &   -8.86  & 19.98 $\pm$ 0.12  \\
    C0337-4  & 0.592    &  1.63 & -1.04  &   850.2 &  11.63  &   -7.43  & 20.03 $\pm$ 0.05  \\
    C0337-5  & 0.589    & -1.49 & -2.44 &   284.7  & -10.66  &   -17.43 & 20.11 $\pm$ 0.05  \\
\end{tabular}
\end{table*}

\section{Models}

The initial conditions for the cluster and the galaxies are
spherically symmetric density distributions. In particular, for the
cluster DM distribution we use a Hernquist (1990) model: note that in
the central regions it is indistinguishable from the Navarro, Frenk \&
White (1996) density profile, while characterized by finite total
mass. Thus,
\begin{equation}
\rhocl (r)= {\Mcltot \rccl \over 2 \pi r (\rccl +r)^{3}},
\end{equation}

\begin{equation}
{\Mcl (r)\over \Mcltot}=\left({r \over \rccl + r}\right)^{2},
\end{equation}

\begin{equation}
{\Psicl (r)}={G \Mcltot \over \rccl + r },
\end{equation}

\noindent
where $\rhocl(r)$ is the cluster density profile, $\Mcl(r)$ and
$\Mcltot$ are the mass within the radius $r$ and the total cluster
mass, respectively, $\Psicl(r)$ is the relative (positive) potential,
and $\rccl$ is the so--called ``core radius''. For the galaxies we use
both one and two--component Hernquist models (Ciotti 1996), and so the
galaxy stellar and DM components are also described by equations
(1)-(3), where the subscripts ``*'' and ``h'' now identify the two
distributions, respectively. In the two--component models we always
assume $\Mhalo\equiv5\Mstar$ and $\rchalo \equiv 3 \rcstar$.

We note that a certain amount of radial orbital anisotropy is expected
in some of the current structure formation scenarios of Es (see, e.g.,
van Albada 1982, Barnes 1992, Hernquist 1993) and clusters (see, e.g.,
Crone, Evrard \& Richstone 1994, Cole \& Lacey 1996, Ghigna et
al. 1998). However, there exist observational and theoretical
indications that the amount of radial anisotropy should be modest in
the both cases (see, e.g., Carollo et al. 1995, Ciotti \& Lanzoni
1997, van der Marel et al. 2000, Gerhard et al. 2001, Nipoti,
Londrillo \& Ciotti 2002, hereafter NLC02), even at significant
look-back times (Koopmans \& Treu 2003). Accordingly, in a subset of
simulations, the cluster and/or the galaxy stellar components are
radially anisotropic, while for sake of simplicity the galactic DM
halo component is always isotropic.  In practice, radial orbital
anisotropy is introduced by using the Osipkov-Merritt parameterization
(Osipkov 1979; Merritt 1985), where the supporting distribution
function (DF) for the density profile $\rho$ is given by
\begin{equation}
f(Q)=\frac{1}{\sqrt{8}\pi^2}\frac{d}{dQ}
       \int_0^Q{\frac{d{\varrho}}{d\psit}}{\frac{d\psit}{\sqrt{Q-\psit}}},
\end{equation}
with
\begin{equation}
\varrho (r)=\left(1+\frac{r^2}{\ra^2}\right)\rho (r),
\end{equation}

\noindent
and $f(Q)=0$ for $Q\leq0$. The quantity $Q$ is defined as $Q\equiv
\en-{L^2/2\ra^2}$, where $\en =\psit-v^2/2$ is the relative (positive)
energy per unit mass, $v$ is the modulus of the velocity vector,
$\psit$ is the relative total gravitational potential, and $L$ is the
modulus of the angular momentum per unit mass.  The parameter $\ra$ is
the so--called ``anisotropy radius''. For $r \gg \ra$ the velocity
dispersion tensor is radially anisotropic, while for $r \ll \ra $ it
is nearly isotropic.  The isotropic DF is given by equation (4) in the
limit $\ra\to\infty$. Following the adopted notation, we indicate the
anisotropy radius of the cluster and of the galaxies as $\racl$ and
$r_{\rm a,*}$, respectively.

\section{Numerical simulations}

\subsection{Initial conditions}

The coordinate system adopted to describe Fig.~1 is defined so that
the $x$-axis runs east--west, the $y$-axis runs south--north, and the
$z$-axis is along the line--of--sight. Observations provide only 3 of
the required 6 initial phase--space coordinates of the center of mass
of each galaxy, namely two positions in the $(x,y)$ projected plane
and the line--of--sight velocity $v_z$. The problem of the orbital
evolution of the five galaxies is thus underdetermined, and the
missing initial coordinates force us towards a {\it probabilistic
approach}, where several initial conditions compatible with the
observational constraints are used to evolve the system from $z=0.59$
to the present.  We assume, for simplicity, that the five galaxies
have equal masses (see Section~2), and that in the inertial reference
system centered on the cluster center
$\sum{x_{i}}=\sum{y_{i}}=\sum{z_{i}}=\sum{v_{x,i}}=\sum{v_{y,i}}=\sum{v_{z,i}}=0$,
summing over the five galaxies. In Table~1 we report the values of the
known coordinates for the five galaxies, reduced to the reference
system (we adopted $\Omega_{\rm m}=0.3$, $\Omega_{\rm \Lambda}=0.7$,
and $H_0=65$ km s$^{-1}$ Mpc$^{-1}$).

In order to fix the missing coordinates $(z_i,v_{x,i},v_{y,i})$ of the
galaxies, as a first step we constrain the structural and dynamical
properties of the cluster model, by choosing the values of the
parameters $\sacl$, $\rccl$, and $\Mcltot$ (see Section 3).  We
investigated two cases for the DF of the cluster: the isotropic case
(corresponding to $\sacl \equiv \racl/\rccl=\infty$) and the radially
anisotropic case with $\sacl=1.8$, a value that fiducially corresponds
to the maximum degree of radial anisotropy compatible with stability
for the one--component Hernquist model (see, e.g., NLC02, and
references therein). For $\rccl$, we adopted the values $100$ kpc and
$300$ kpc (corresponding to half-mass radius $r_{\rm M,cl} \simeq 241$
kpc and $r_{\rm M,cl} \simeq 724$ kpc, respectively): note that, at
least in projection, all the five galaxies are well within $\rccl$.
Obviously, for a fixed $\rccl$ there is a minimum cluster mass
($\Mmin$) for which all the five galaxies are bound, under the
implicit assumption that the cluster DM is virialized (consistently
with the time independence of the cluster potential)\footnote{This
condition will be relaxed in a set of 4 simulations described in
Section~6, in which we study the evolution of the 5 galaxies in a {\it
collapsing} cluster DM distribution. Also, in Section~6 we present a
simulation in which a population of small galaxies is added to the
cluster.}. For each galaxy, the lower limit to the
cluster mass corresponds to the case of vanishing $v_x$, $v_y$ and
$z$: then $\Mmin$ is given by the maximum of these 5 lower
limits. From the values in Table~1 and from equation (3), $\Mmin
\simeq 4.4\times10^{13}\Msol$, when $\rccl=100$ kpc, and $\Mmin \simeq
1.28\times10^{14}\Msol$, in the case $\rccl=300$ kpc. It is clear that
in the limiting case $\Mcltot=\Mmin$, the least bound of the galaxies
has a vanishing phase--space volume available: for this reason, in the
numerical simulations, for each choice of $\rccl$ and $\sacl$, we
explore the cases $\Mcltot\,\gsim\,\Mmin$ and $\Mcltot\sim2\Mmin$. In
the first case one of the galaxies is weakly bound, while in the
second case all the galaxies are expected to be well bound. The exact
values of $\Mcltot$ are reported, for each simulation, in Table~2.

Now that the properties of the cluster are fixed, we can use general
physical principles to constrain the missing phase--space
information. A first basic requirement that we impose on the unknown
coordinates $(z,v_x,v_y)$ of each galaxy is that they correspond to
objects bound to the cluster, i.e., for each galaxy $v_x^2+v_y^2<2
\Psicl (x,y,z)-v_z^2$, where we neglected the galaxy--to--galaxy
contribution to the binding energy.  In principle, for a given cluster
density profile, and without extra assumptions on the dynamical status
of the five galaxies, all the sets of phase--space coordinates
corresponding to bound galaxies and to a null barycentric motion
should be accepted. Under the hypothesis that the five galaxies follow
the same DF as the cluster DM , we can proceed with a more detailed
discussion on the selection of the initial conditions.

In our approach, we first obtain the coordinate $z$ for each galaxy,
by applying the von Neumann rejection method (see, e.g., Aarseth,
Henon \& Wielen 1974) to the mass profile of the cluster. Once the
position of the galactic center of mass is fixed, we recover the two
unknown velocities $v_x$ and $v_y$, again by application of the von
Neumann rejection method to the cluster DF, where $r$ and $v_z$ are
fixed.  As a rule, when extracting the initial conditions, we discard
the realizations in which the barycenter position of the group of the
5 galaxies deviates from the cluster center more than $0.1\rccl$
and/or its velocity is larger than $0.1(G\Mcltot/\rccl)^{1/2}$.  We
also performed a few simulations in which the barycentric property of
the five galaxies is perfectly realized, finding that the results of
interest are not affected by the assumed tolerance on the center of
mass of the system.

At the beginning of each simulation the five galaxies are identical
Hernquist models with core radius $\rcstar \simeq 2.2$ kpc (which
corresponds to an effective radius $\Re\simeq4$ kpc). In the one--component
case the total stellar mass of each galaxy is $\Mstar=4 \times 10^{11}
\Msol$, while in the two--component case is reduced to $\Mstar=2
\times 10^{11} \Msol$, and the galactic DM halo is more massive and
more extended than the stellar component ($\Mhalo=10^{12} \Msol$ and
$\rchalo\simeq6.6$ kpc). When using two--component galaxy models, we
reduce the diffuse dark component of the cluster by an amount
corresponding to $5\Mhalo$, so that the total amount of DM in the
N-body simulations is the same as in the one--component case.  We also
explore some cases of (one--component) radially anisotropic galaxy
models, by assuming $\sast \equiv r_{\rm a,*} / \rcstar=1.8$. We
define the half--mass dynamical time of the galaxies as
\begin{equation}
\Td \equiv \sqrt{\frac{3\pi}{16G\rhoM}},
\end{equation}
where $\rhoM =3(\Mstar+\Mhalo)/8\pi \rme^3$ is the mean density inside
the half--mass radius $\rme$ of the total (stellar plus DM)
distribution. With the adopted values of the parameters, the
half--mass dynamical time of the galaxies is $\Td \simeq 2.0 \times
10^{7}$~yr and $\Td \simeq 4.8 \times 10^{7}$~yr in the one and
two--component case, respectively.

Summarizing, our initial conditions are characterized by the
properties of the cluster, which are fully determined by the three
parameters ($\rccl,\Mcltot,\sacl$), by those of the galaxies (presence
or absence of galactic DM halos, $\sast$), and by the particular
realization considered. Clearly, the results of the simulations
depend, for a given set of cluster and galaxies parameters, also on
the specific values of the initial positions and velocities of the
five galaxies. Thus, it is natural to wonder about the statistical
significance that should be associated to the result of a single
simulation or to a set of simulations relative to given cluster and
galaxies parameters.

In order to address this issue, we start by considering the idealized
case in which the whole available parameter space is explored by the
simulations. In this case, any result of interest $\res$ (e.g., the
number of merging galaxies or the time of the last merging) is a
function of the missing phase--space coordinates: $\res=\res({\bf
w_1,w_2,w_3,w_4,w_5})$, where ${\bf w}_i=(z_i,v_{x,i},v_{y,i})$ for
$i=1,5$.  Under the additional assumption that the dynamical evolution
of each galaxy is independent of the initial positions and velocities
of the other galaxies (justified in the considered scenario, in which
the dominant dynamical mechanism is the dynamical friction of the
galaxies against the diffuse cluster DM), the statistical weight of
each simulation can be obtained by considering the product of the five
reduced DFs, $f_i({\bf w}_i) = f(x_{i,0},y_{i,0},v_{z,i,0},{\bf
w}_i)$. Accordingly, the statistically weighted result can be written
${\langle {\res}\rangle} = {{\mathcal{N}}^{-1}} \int {\res} ({\bf
w}_1,{\bf w}_2,{\bf w}_3,{\bf w}_4,{\bf w}_5) \prod_{i=1}^5 f_i{(\bf
w}_i) d^3{\bf w}_i $, where the normalization ${\mathcal{N}}$ is given
by ${\mathcal{N}} = \int \prod_{i=1}^5 f_i{(\bf w}_i) d^3{\bf w}_i$.
In case of a finite number of simulations $N$, the previous relations
become
\begin{equation}
{\langle{\res}\rangle} = {1 \over {\mathcal{N}}} \sum_{k=1}^N {\res}({\bf w}_{1,k},{\bf w}_{2,k},{\bf w}_{3,k},{\bf w}_{4,k},{\bf w}_{5,k}) \prod_{i=1}^5 f_i{(\bf w}_{i,k}),
\end{equation}
where now  
\begin{equation}
{\mathcal{N}} =  \sum_{k=1}^N \prod_{i=1}^5 f_i{(\bf w}_{i,k}).
\end{equation}
In practice, we considered as a rule just 2 realizations for each set
of parameters (see Table~2), while in one case we explored 7 different
realizations (simulations~\#7-13). In any case, we will use equations
(7) and (8) in order to quantify the expected merging time.

\subsection{Numerical methods}

For the numerical N-body simulations we used both the serial and
parallel versions of the Springel, Yoshida \& White (2001) GADGET
code. Once the position and velocity of the center of mass of each
galaxy were fixed by using the approach described in Section 4.1, the
numerical realization of the initial conditions for the galaxies and
for the cluster DM distribution was obtained by following the scheme
described in NLC02.

For the purpose of this work we are interested in the dynamical
evolution of the system up to $z=0$. Thus, the total time of each
simulation is $t_{\rm tot}=t(0)-t(z_{\rm cl})$, where $z_{\rm cl}
= 0.59$ is the redshift of the cluster.  In the adopted standard
$\Lambda{\rm CDM}$ cosmology (see Section 4.1) $t_{\rm tot}\simeq 6.1$
Gyr, a time of the order of $100\,\Td$ for the galaxies (see equation
6) and of $10 - 50\,\Td$ for the cluster (depending on $\Mcltot$ and
$\rccl$).  All the relevant properties of the numerical simulations
are reported in Table~2, where the results within each group
correspond to different realizations of the initial conditions for the
same cluster parameters.

The choice of the number of particles was determined by computational
time limits and by the requirement that all the particles (DM and
``stars'') have the same mass. For these reasons, simulations
characterized by different cluster and galaxy parameters were run with
different number of particles.  We note that in case of high mass
ratio between the cluster and the galaxies (for example, when
$\rccl=300$ kpc and $\Mcltot\sim2\Mmin$) even a quite large number of
cluster particles ($\Ncl\sim2\times10^5$) implies a small number of
stellar galaxy particles ($\Nstar=256$).

Five parameters characterize GADGET simulations: the cell--opening
parameter $\alpha$, the minimum and the maximum time step $\Dtmin$ and
$\Dtmax$, the time step tolerance parameter $\alphatol$, and the
softening parameter $\varepsilon$ (Springel et al. 2001).  We adopted
$\alpha=0.02$, $\Dtmin =0$, $\Dtmax =\Td/100$ (where $\Td$ is
evaluated for the initial conditions of the galaxies),
$\alphatol=0.05$, and $\varepsilon=\Re/5\simeq0.36 \rcstar$ (where
$\Re$ is the initial effective radius of the galaxies). With these
choices we obtained a conservation of the total energy with deviations
that do not exceed $|\Delta E/E |\simeq 1 \%$ over the entire
simulation.

\section{Results}

\begin{table*}
 \centering
  \caption{Simulations parameters.}
  \begin{tabular}{l|rrrr|crr|cr|cr}
\# &$\rccl$&$\Mcltot$&$\sacl$&$\Ncl$&$\Mstar$&$\sast$&$\Nstar$&$\Mhalo/\Mstar$&$\Nhalo$&$\Nmerg$&$\Tlast$ \\
\hline
\hline
1  &  100     & 4.8            & 1.8      & 235520       & 4               & $\infty$& 2048  & 0 & -  &   4           &      3.0     \\
2 &  100     & 4.8            & 1.8      &  29440       & 4               & $\infty$&  256  & 0 & -  &   4           &      1.0     \\
1a &  100     & 4.8            & 1.8      &  58880       & 4               & 1.8     &  512  & 0 & -  &   4           &      3.0     \\
1h &  100     & 4.8            & 1.8      & 107520       & 2               & $\infty$&  512  & 5 & 2560 &   4           &      2.0     \\
\hline 
3  &  100     & 5.3            & $\infty$ & 130560       & 4               & $\infty$& 1024  & 0 & - &   4           &      1.5    \\
4  &  100     & 5.3            & $\infty$ &  32640       & 4               & $\infty$&  256  & 0 & - &   4           &      2.0    \\
3a  &  100     & 5.3            & $\infty$ &  65280       & 4               & 1.8     &  512  & 0 & -  &   4           &      1.5      \\
3h  &  100     & 5.3            & $\infty$ & 120320       & 2               & $\infty$&  512  & 5 & 2560  &   4           &      1.0    \\
\hline 
5 &  100     & 9.6            & 1.8      &  60160       & 4               & $\infty$&  256  & 0 & -   &   5           &      3.0    \\
6 &  100     & 9.6            & 1.8      & 120320       & 4               & $\infty$&  512  & 0 & -   &   5           &      3.5     \\
5a &  100     & 9.6            & 1.8      & 120320       & 4               & 1.8     &  512  & 0 & -  &   5           &      3.0    \\
6h &  100     & 9.6            & 1.8      & 115200       & 2               & $\infty$&  256  & 5 & 1280 &   5           &      3.5   \\
\hline 
7  &  100     & 10.6           & $\infty$ & 266240       & 4               & $\infty$& 1024  & 0 & -   &   5           &      2.5      \\
8  &  100     & 10.6           & $\infty$ &  66560       & 4               & $\infty$&  256  & 0 & -   &   5           &      4.5      \\
9  &  100     & 10.6           & $\infty$ &  66560       & 4               & $\infty$&  256  & 0 & -   &   5           &      3.5      \\
10  &  100     & 10.6           & $\infty$ &  66560       & 4               & $\infty$&  256  & 0 & -   &   5           &      4.0      \\
11  &  100     & 10.6           & $\infty$ &  66560       & 4               & $\infty$&  256  & 0 & -   &   5           &      2.0      \\
12  &  100     & 10.6           & $\infty$ &  66560       & 4               & $\infty$&  256  & 0 & -   &   5           &      2.5      \\
13  &  100     & 10.6           & $\infty$ &  33280       & 4               & $\infty$&  128  & 0 & -   &   5           &      2.5      \\
7a  &  100     & 10.6           & $\infty$ & 133120       & 4              & 1.8     &  512  & 0 & -  &   5            &      2.5   \\
7h  &  100     & 10.6           & $\infty$ & 128000       & 2              & $\infty$&  256  & 5 & 1280  &   5         &      2.5    \\
\hline 
14 &  300     & 13.5           & 1.8      &  85120       & 4               & $\infty$&  256  & 0 & -   &   4           &      1.5     \\
15 &  300     & 13.5           & 1.8      &  85120       & 4               & $\infty$&  256  & 0 & -  &   4            &      2.5     \\

14a &  300     & 13.5           & 1.8      & 170240       & 4               & 1.8     &  512  & 0 & -  &   4           &      1.5     \\
14h &  300     & 13.5           & 1.8      & 165120       & 2               & $\infty$&  256  & 5 & 1280 & 4           &      1.5    \\
\hline 
16 &  300     & 15.3           & $\infty$ &  96640       & 4               & $\infty$&  256  & 0 & -  &   4            &      2.5      \\
17 &  300     & 15.3           & $\infty$ & 193280       & 4               & $\infty$&  512  & 0 & -  &   3            &      2.0      \\
16a &  300     & 15.3           & $\infty$ &  96640       & 4               & 1.8     &  256  & 0 & -   &   4          &      2.5     \\
17h &  300     & 15.3           & $\infty$ & 188160       & 2               & $\infty$&  256  & 5 & 1280  &   5        &      5.5    \\
\hline 
18 &  300     & 27.0           & 1.8      & 171520       & 4               & $\infty$&  256  & 0 & -   &   4           &      3.0     \\
19 &  300     & 27.0           & 1.8      & 171520       & 4               & $\infty$&  256  & 0 & -  &    5           &      6.0    \\
18a &  300     & 27.0           & 1.8      & 171520       & 4               & 1.8     &  256  & 0 & -  &   4           &      3.0      \\
18h &  300     & 27.0           & 1.8      & 168960       & 2               & $\infty$&  128  & 5 & 640  &   5         &      5.5   \\
\hline 
20 &  300     & 30.6           & $\infty$ & 194560       & 4               & $\infty$&  256  & 0 & -  &   3           &      1.0     \\
21 &  300     & 30.6           & $\infty$ & 194560       & 4               & $\infty$&  256  & 0 & -   &   4           &      1.5    \\

20a &  300     & 30.6           & $\infty$ & 194560       & 4               & 1.8     &  256  & 0 & -  &   3           &      1.0     \\
20h &  300     & 30.6           & $\infty$ & 192000       & 2               & $\infty$&  128  & 5 & 640  &   3           &      1.0    \\
\hline \hline
3c  &  100     & 5.3            & $\infty$ & 130560       & 4               & $\infty$& 1024  & 0 & - &   4           &      1.5    \\
3cc  &  100     & 5.3            & $\infty$ & 130560       & 4               & $\infty$& 1024  & 0 & - &   5           &      6.0    \\
17c &  300     & 15.3           & $\infty$ & 193280       & 4               & $\infty$&  512  & 0 & -  &   4            &      5.0    \\
17cc &  300     & 15.3           & $\infty$ & 193280       & 4               & $\infty$&  512  & 0 & -  &   4            &      5.0    \\
\hline \hline
1s  &  100     & 4.8            & 1.8      & 235520       & 4               & $\infty$& 2048  & 0 & -  &   4           &      2.5     \\
\hline \hline
1f &  100     & 4.8            & -        & -$\quad$      & 4               & $\infty$& 2048  & 0 & -     &   2       &      1.5      \\
7f &  100     & 10.6           & -       & -$\quad$          & 4               & $\infty$& 1024   & 0 & -  &   2           &      0.5      \\
\hline \hline
\end{tabular}

\medskip

\flushleft{ First column: name of the simulation. $\rccl$:~cluster
core radius in kpc. $\Mcltot$:~cluster mass in units of
$10^{13}\Msol$. $\sacl$:~cluster anisotropy parameter.  $\Ncl$:~number
of cluster particles. $\Mstar$:~galaxy stellar mass in units of
$10^{11}\Msol$. $\sast$:~galaxy anisotropy parameter. $\Nstar$:~number
of stellar particles per galaxy. $\Mhalo$: galaxy halo mass in units
of $10^{11}\Msol$. $\Nhalo$:~number of halo particles per
galaxy. $\Nmerg$:~number of merging galaxies. $\Tlast$: time elapsed
from the beginning of the simulation when the last merging occurs (in
Gyrs).  The subscript ``a'' to the simulation name indicates that the
galaxies are anisotropic, while ``h'' indicates the presence of
galactic DM halos; ``c'' and ``cc'' mean that the cluster DM
distribution is collapsing, with initial virial ratio $2T/|W|=0.8$ and
$2T/|W|=0.5$, respectively; the subscript ``s'' indicates that a
population of small galaxies is added to the diffuse DM to represent
the cluster; finally, ``f'' means that the cluster potential is
maintained fixed during the simulation.}

\end{table*}

\subsection{Merging statistics and time-scales}

\begin{figure}
\begin{center}
\parbox{1cm}{ \psfig{file=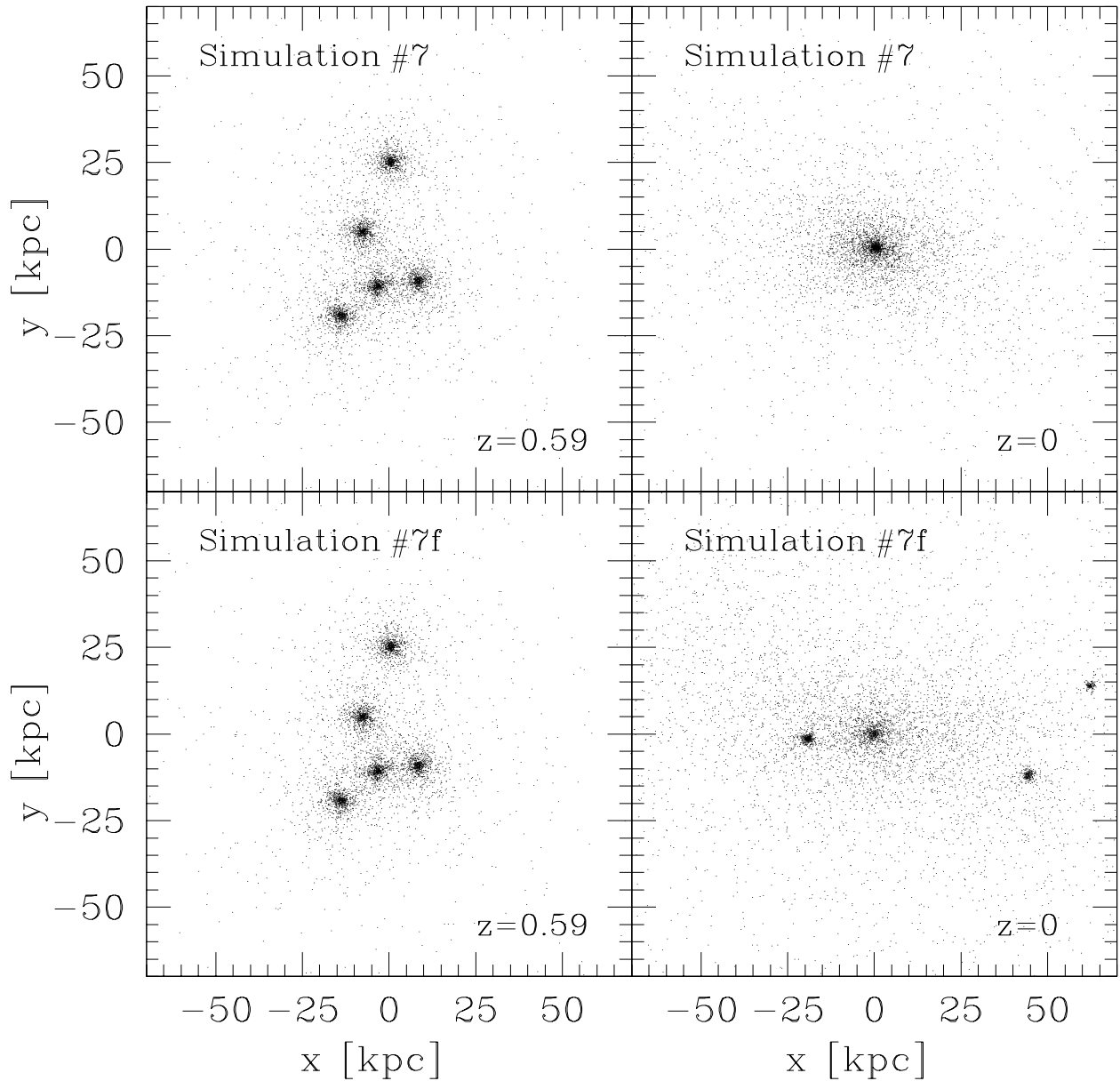,height=17cm}}
\caption{ {\it Top panels}: snapshots of the initial ($z=0.59$, left
panel) and final ($z=0$, right panel) distribution of the stellar
particles in the projected plane ($x$,$y$) for simulation
\#7. {\it Bottom panels}: the same as top panels, but for 
simulation~\#7f. We recall here that the two simulations have
identical initial conditions, the only difference being that in 
simulation~\#7 the cluster DM is represented with particles, while in
simulation~\#7f as a fixed potential (see Table~2). The effect of
dynamical friction is apparent when comparing the final snapshots.}
\label{appenfig}
\end{center}
\end{figure}

The first goal of this work is to investigate whether any, a few, or
all of the five galaxies will have merged into a unique system before
$z=0$, to verify whether the studied system is indeed a good candidate
to represent a case of galactic cannibalism. In Column~11 of Table~2
we report the number $\Nmerg$ of galaxies involved in a merging within
the total time of the simulation (6.1 Gyr), and, in Column 12, the
time $\Tlast$ at which the last merging occurs (calculated from the
beginning of the simulation, with a resolution of 0.5 Gyr). A first
inspection of Table~2 (leaving out the ``special'' simulations~\#1f,
\#7f, \#3c,cc, \#17c,cc, and \#1s) reveals that, in {\it all} the
performed simulations, at least 3 galaxies merge before $z=0$, thus
suggesting that {\it a multiple merging event in the central group of
five galaxies in the cluster C0337-2522 will take place in the next
few Gyrs}.

As already pointed out in the Introduction, in our investigation we
considered the possibility that the driving mechanism leading to
merging is the dynamical friction of the galaxies against the cluster
DM, making them spiral towards the cluster center. In order to test
this hypothesis, we also ran two simulations by modeling the cluster
as a {\it frozen} DM distribution: these two simulations (\#1f and
\#7f) have the same initial conditions as simulations~\#1 and \#7,
respectively. We found that, at variance with simulations~\#1 and \#7,
in simulations~\#1f and \#7f only 2 galaxies merge before $z=0$, and
on the basis of these results we confirm that the dynamical friction
of the galaxies against the cluster DM is the primary mechanism
responsible for the galactic cannibalism.  The merging of two galaxies
in case of frozen halo can thus be interpreted as a result of the less
important effect of galaxy--galaxy interaction. In Fig.~2 we plot, as
an example, the initial (left panels) and final (right panels)
distributions in the projected plane ($x$,$y$) of the {\it stellar}
particles for simulations~\#7 and \#7f: at $z=0$ a single galaxy is
formed in case of live cluster DM (upper right panel in Fig.~2), while
four distinct stellar systems are still present in case of frozen
cluster DM (lower right panel in Fig.~2).

As already pointed out in Section 4.1, for fixed cluster parameters we
explored, as a rule, 2 (but in one case 7) different realizations of
the initial conditions, identified in Table~2 by groups separated by
horizontal lines. One of the realizations in each group is also used
as initial condition for simulations with anisotropic galaxy models
(named in Table~2 with the number of the corresponding isotropic
simulation with the subscript ``a''); similarly, in order to explore
the effects of the presence of galactic DM halos, for each group we
ran also a simulation with two--component galaxy models (named in
Table~2 with the number of the corresponding one--component simulation
with the subscript ``h''). As expected, the number of merging galaxies
$\Nmerg$ and the characteristic merging time-scale do not change if
anisotropic galaxy models are used in the initial conditions.  We
recall here that the choice of exploring cases with anisotropic
initial galaxy models was aimed at investigating possible effects on
the properties of the end--products (see following Sections).  As also
expected, in the two--component simulations (in which by construction
more massive galaxy models are used) dynamical friction is more
effective than in absence of galactic DM. In some of the
two--component cases more galaxies merge than in the corresponding
one--component simulations, while in others $\Nmerg$ is the same, but
$\Tlast$ is shorter (of $0.5-1$ Gyr). We note that $\Tlast$ is by
definition dependent on the number of merging and it is not a direct
measure of the dynamical friction time-scale ($\Tfric$, that we define
empirically as the time in which a galaxy reaches the center of the
cluster as a consequence of the interaction with the diffuse DM). This
can be seen, for example, by considering simulations~\#3 and \#7. In
the former, 4 galaxies merge in 1.5~Gyr, while, in the latter, 5
galaxies merge in 2.5~Gyr. In this case, the dynamical friction
time-scale, as defined above, is of course shorter in the case of 5
merging galaxies, even if $\Tlast$ is larger.

A more detailed analysis is required to address the dependence of the
number of merging galaxies, and of the merging time-scales, on the
cluster parameters and on the particular realization considered.  We
found that $\Tlast$ depends on both the cluster parameters and the
realization. This is not surprising, since the dynamical friction
time-scale is a function of both the cluster density and the initial
velocity of the galaxies.  The general trend is that the number of
merging galaxies is nearly independent of the specific realization for
given cluster parameters, and does not depend strongly on the cluster
properties either. In general, $\Nmerg$ is found to be insensitive to
the adoption of the ``minimum mass'' hypothesis ($\Mcltot\, \gsim\,
\Mmin$ instead of $\Mcltot\sim 2 \Mmin$). However, one could ask what
is special with the set of simulations from \#5 to \#13, in all of
which 5 galaxies merge.  In order to answer this question it is
necessary to try a rough quantitative evaluation of the dynamical
friction time-scale $\Tfric$.  As it is well known, $\Tfric \propto
v^3 / \rhocl$, where $v$ is the modulus of the velocity vector of the
galaxy and $\rhocl$ is the cluster density (see, e.g., Binney \&
Tremaine 1987): roughly $\rhocl\sim\Mcltot/\rccl^3$, and so $\Tfric
\propto v^3 \rccl^3 / \Mcltot$. Thus, it is clear that {\it for fixed
galaxy velocity} the dynamical friction time-scale decreases for
decreasing cluster radius and for increasing cluster mass, and the
factor $\rccl^3 / \Mcltot$ is minimized, in our exploration, for
cluster parameters of simulations from \#5 to \#13.  {\it We also note
that in the cases in which 4 galaxies merge it is always galaxy
C0337-2 (characterized at $t=0$ by the highest absolute value of
line--of--sight velocity, see Table~1) that survives as an individual
object for the time interval covered by the simulations: a clear
consequence of its high velocity}.

Finally, on the basis of the discussion at the end of Section~4.1., we
can determine, for each set of simulations with the same cluster and
galaxy properties, the statistically weighted value of the merging
time-scale. As an example, we focus here on the set of 7 simulations
from \#7 to \#13, in all of which 5 galaxies merge.  By applying
equation (7), considering as result of interest the time of last
merging $\Tlast$, we find that in this case the statistically weighted
last merging time is ${\langle{\Tlast}\rangle}\simeq2.3\pm0.3$ Gyrs,
where the uncertainty has been computed by assuming an uncertainty of
$0.5$ Gyr associated to $\Tlast$ in each simulation.

\subsection{Properties of the end--products}

We define end--product of a simulation the stellar system composed by
the bound particles initially belonging to the galaxies involved in
the merging process. In evaluating the binding energy of the
particles, we consider the gravitational potential of both the cluster
and the remnant galaxy mass distribution.  We found that the fraction
of unbound particles is negligible (in any case smaller than 0.2 per
cent), and thus the mass of the remnant is given by the sum of the
masses of its progenitors.

We measured some intrinsic and projected quantities of the
end--products: the intrinsic axis ratios $c/a$ and $b/a$ (where
$a,b,c$ are respectively the longest, intermediate and shortest axis
of the associated inertia tensor), the angle--averaged half mass
radius, the virial velocity dispersion and the total angular momentum,
and, for a set of 50 random projections, the circularized effective
radius, the projected central velocity dispersion, the ellipticity and
the circularized SB profile.  In the treatment of the outputs of the
numerical simulations we followed the scheme described in NLC02, and,
in order to limit the uncertainties due to discreteness effects, we
analysed the intrinsic and ``observational'' properties of the
end--products with at least $n \times 512$ particles (where $n$ is the
number of galaxies involved in merging). To satisfy this condition and
to explore the properties of the end--products of all the considered
initial conditions, we ran, for each choice of the cluster parameters,
at least one simulation with 512 stellar particles per galaxy. The
only exception is the case ($\rccl=300$ kpc, $\Mcltot\sim2\Mmin$) in
which we used 256 stellar particles per galaxy.

\subsubsection{Structural and dynamical parameters}

The end--product is in general well described by a triaxial ellipsoid
with axis ratios in the range $0.5\,\lsim\,c/a\,\lsim\,0.9$. In a few
cases we found oblate systems with $c/a \simeq c/b \simeq 0.5 \, - \,
0.7$. These oblate systems are mainly flattened by rotation: their
angular momenta (normalized to the typical scales of the system: total
mass $M$, virial velocity dispersion $\sgv$, and half mass radius
$\rme$) are, in modulus, among the highest observed in the sample. In
addition, a significant degree of alignment between the angular
momentum and the minor axis of the inertia tensor is observed in these
cases.

In the case of isotropic one--component initial models, the
end--products of merging of 5 galaxies (with total stellar mass
$\Mstar\simeq 2\, \times 10^{12} \Msol$) have circularized effective
radius in the range $12\, \lsim\, \cRe\, \lsim\, 21$ kpc, and central
velocity dispersion (measured inside an aperture of equivalent radius
$\cRe/4$)\footnote{Note that the simulated aperture is of the order of
the adopted softening length $\varepsilon$ (see Section 4.2). We ran a
subset of simulations with $\varepsilon/4$: while the number of
merging galaxies and the merging times resulted unaffected, $\sg0$
resulted increased at most by a factor of $1.15$, less than the
observational scatter and/or projection effects in Figs~4,~5,~6.} in
the range $320\, \lsim\, \sg0 \, \lsim\, 410 \, {\rm km \, s^{-1}}$.
In case of merging of 3 or 4 galaxies ($\Mstar\simeq 1.2 \times
10^{12} \Msol$ and $\Mstar\simeq 1.6 \times 10^{12} \Msol$,
respectively) we found $10\,\lsim\, \cRe\, \lsim\, 12$ kpc, and $290\,
\lsim\, \sg0 \, \lsim\, 392 \, {\rm km \, s^{-1}}$; approximately the
same ranges are spanned by $\cRe$ and $\sg0$ of the end--products of
merging of radially anisotropic galaxy models. The main
characteristics of the end--products of two--component galaxies are
not substantially different from those of the corresponding
one--component cases, with the only exception of the central velocity
dispersion, which is in general quite high ($300\, \lsim\, \sg0 \,
\lsim\, 491 \, {\rm km \, s^{-1}}$), while as a rule $7\, \lsim\,
\cRe\, \lsim\, 17$ kpc.

Thus, the values of $\cRe$ are always comparable with those measured
in real luminous Es (see, e.g., J{\o}rgensen, Franx \& Kj{\ae}rgaard
1996). In case of one--component progenitors, also $\sg0$ lies in the
same range as those measured in observations, while in a few
simulations with two--component galaxies, the remnant is characterized
by very large values of $\sg0$, unusual even for giant Es and cD
galaxies in the center of clusters, which have $\sg0 \,\lsim\, 400\,
{\rm km \, s^{-1}}$ (see, e.g., Oegerle \& Hoessel 1991).

\begin{figure}
\begin{center}
\parbox{1cm}{ \psfig{file=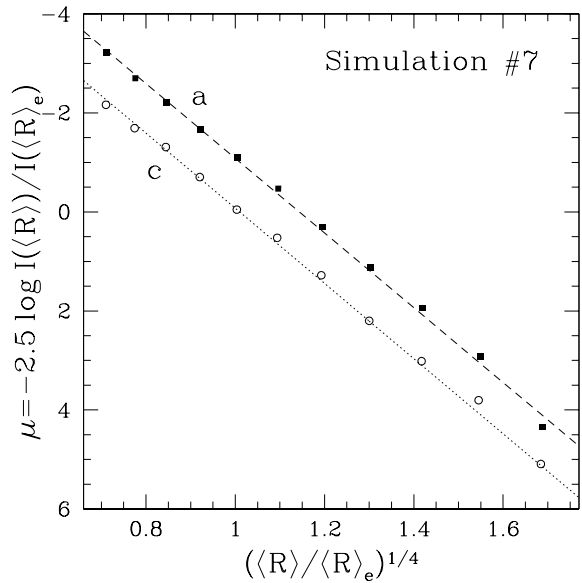,width=8cm}}
\caption{Circularized SB profiles of the end--product of 5 merging
galaxies (simulation~\#7) projected along its major axis (a; solid
squares) and minor axis (c; empty circles). The dashed and dotted
curves are the corresponding $R^{1/4}$ best--fits. For clarity, an
artificial vertical shift has been applied to the first profile.}
\label{appenfig}
\end{center}
\end{figure}

\subsubsection{Surface brightness profiles}

One of the motivations for this work was to test the hypothesis that
the system of five galaxies under investigation could be considered
the progenitor of a cD galaxy. The most recognizable feature of cD
galaxies is their SB profile, characterized by the $R^{1/4}$ law in
the inner part and by a systematic deviation from this law in the
outer part (roughly for ${\ciR/\cRe\, \gsim\, 3}$, where ${\langle
R\rangle}$ is the circularized projected radius), due to the presence
of a diffuse luminous halo (see, e.g., Sarazin 1986, Tonry 1987). We
analysed the circularized SB profiles of three different projections
(along the three principal axes) of each end--product, by fitting them
with the standard de Vaucouleurs (1948) $R^{1/4}$ law over the radial
range $0.2\,\lsim\, \ciR/\cRe\,\lsim\, 10$: overall, the $R^{1/4}$
fits can be considered in good agreement with the profiles (see, e.g.,
Fig.~3, where we plot the circularized SB profiles of the end--product
of simulation~\#7), even though the average residuals between the data
and the fits were found in the range $0.2 \, \lsim \, \Dmu \, \lsim \,
0.5\,({\rm mag}\,{\rm arcsec^{-2}})$.  These residuals are not small,
but the deviation from the $R^{1/4}$ law is not a systematic excess at
large radii, as can be seen from Fig.~3. We also fitted the profiles
with the Sersic (1968) $R^{1/m}$ law:
\begin{equation}
I(R)=I_0\,\exp\left[-b(m)\left(\frac{R}{\Re}\right)^{1/m}\right],
\end{equation}
where $b(m) \simeq 2m-1/3+4/(405m)$ (Ciotti \& Bertin 1999).  Thanks
to the additional parameter $m$, we obtained better fits of the SB
profiles with the best--fitting parameter $m$ in the range $3.5 \,
\lsim\, m \,\lsim\, 6.8$ and $0.05 \, \lsim \, \Dmu \, \lsim \, 0.25
\,({\rm mag}\,{\rm arcsec^{-2}})$, while for the Hernquist profile of
the initial galaxies we found $m\simeq 3.5$, always over the radial
range $0.2\,\lsim\, \ciR/\cRe\,\lsim\, 10$. Thus, the trend is of $m$
increasing with merging, in agreement with what found by Londrillo,
Nipoti \& Ciotti 2003 and Nipoti, Londrillo \& Ciotti (2003a,
hereafter NLC03a), who consider higher resolution N-body simulations
of merging hierarchies.  Also the end--products obtained from merging
of anisotropic initial systems have SB profiles fitted quite well by
the $R^{1/4}$ law up to $\ciR\simeq 10 \cRe$, with average residuals
$0.1 \, \lsim \, \Dmu \lsim \, 0.6 \, ({\rm mag}\,{\rm
arcsec^{-2}})$. Adopting the Sersic law as fitting function, we find
the best--fitting parameter in the range $3.7 \, \lsim\, m \, \lsim \,
6.4$.  In addition, there is no significant difference in the light
distribution of one and two--component end--products: also in the
presence of galactic DM halos, there is no evidence of any systematic
excess at large radii in the SB profile.

On the basis of these results, {\it we find no indications that the
merger remnant will be similar to a cD galaxy.  In contrast, it seems
that the product of a multiple merging like that considered is more
similar to a ``normal'' giant elliptical}. This is in agreement with
Zhang et al. (2002) who have shown, for a different set of initial
conditions, that collapse with substructure is unable to produce a cD
halo (see also Section~6).

\begin{figure}
\begin{center}
\parbox{1cm}{ \psfig{file=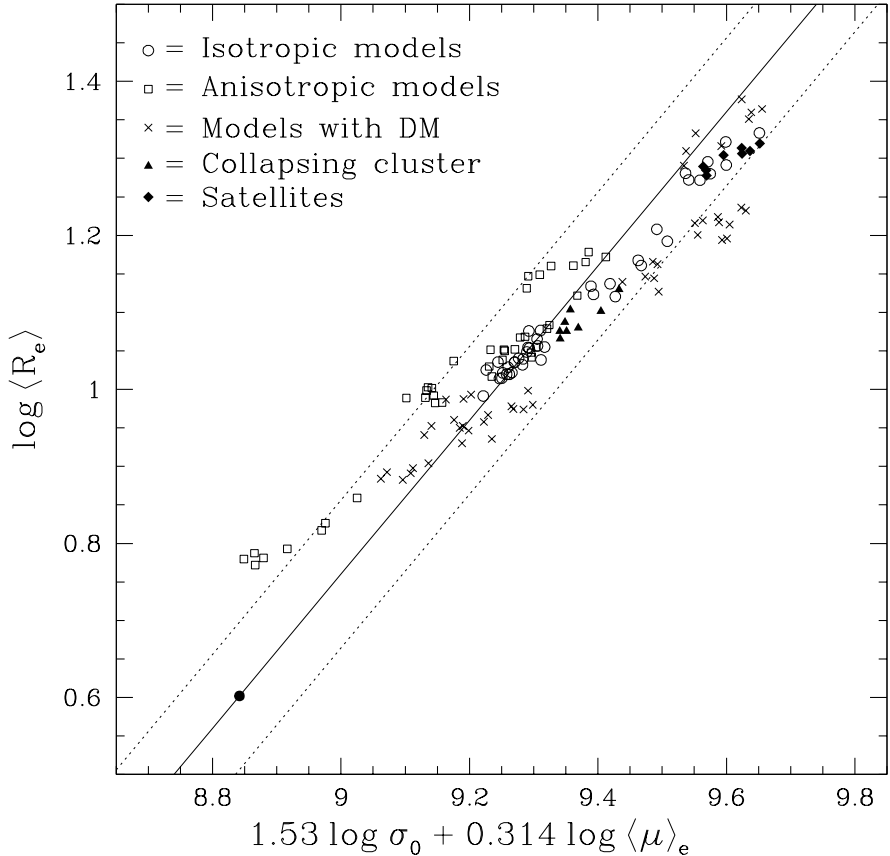,height=12cm}}
\caption{K-band FP: observational best--fit relation (solid line; Pahre
et al. 1998) and its scatter (dotted lines). Filled circle: initial
conditions. Empty circles and squares: end--products of the merging of
isotropic and anisotropic one--component galaxy models, respectively
(simulations \#1, 3, 6, 7, 17, 1a, 3a, 5a, 7a, 14a). Crosses:
end--products in the two--component case (simulations \#1h, 3h, 6h,
7h, 14h, 17h). Triangles and diamonds: merger remnants of simulations
\#3c and \#1s, respectively. The plot represents 8 random projections
of each end--product.}
\label{appenfig}
\end{center}
\end{figure}

\begin{figure}
\begin{center}
\parbox{1cm}{ \psfig{file=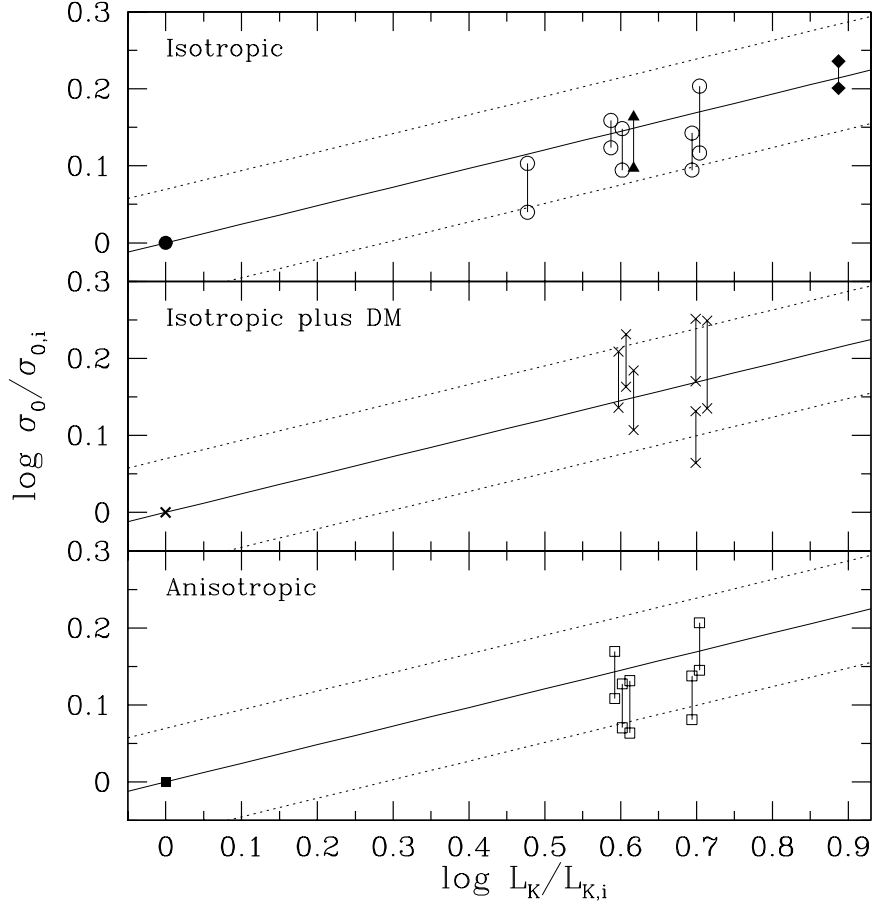,height=12cm}}
\caption{Central velocity dispersion of the final system vs. its total
K-band luminosity in case of isotropic progenitors, progenitors with
DM halo, and anisotropic progenitors (for the same subset of
simulations as in Fig.~4 and adopting the same symbols). $L_{\rm K,i}$
and $\sigma_{\rm 0,i}$ are the common luminosity and central velocity
dispersion of each progenitor. The solid and dotted lines represent
the observed near-infrared FJ relation with its r.m.s. scatter,
respectively.  Vertical bars span the whole range of values of $\sg0$
due to the orientation of the line--of--sight with respect to each
end--product: the horizontal split in each group of vertical bars is
artificial, and it is introduced for clarity. }
\label{appenfig}
\end{center}
\end{figure}

\subsubsection{Fundamental Plane and Faber--Jackson relations}

The previous analysis showed that the merging end--products have SB
profiles more similar to Es than to cDs. However, as briefly discussed
in the Introduction, real Es follow well defined scaling relations.
Thus, it is of particular interest to investigate whether the
end--products satisfy the FP and the FJ relations. We consider the FP
relation in the near-infrared K-band, with observational best--fit
\begin{equation}
{\rm log}\,\cRe=1.53\,{\rm log}\,\sg0+0.314 \, \mue - 8.24,
\end{equation}
where $\mue=-2.5\,{\rm log}\,\Lk/2\pi{\cRe}^2$ is expressed in mag
arcsec$^{-2}$, $\sg0$ in km s$^{-1}$ and $\cRe$ in kpc ($\Lk$ is the
total luminosity in the K-band, the additive constant is evaluated for
$h=0.65$, and the scatter of ${\rm log}\,\cRe$ around the best--fit
has r.m.s.=0.096; Pahre, Djorgovski \& de Carvalho 1998).  In the
following we also consider the K-band FJ relation given by Pahre et
al. (1998):
\begin{equation}
\Lk  \propto  \sg0^{4.14}, 
\end{equation}
with a reported scatter of $0.72$ mag (for Coma cluster Es).

The FP and FJ relations are known to evolve with redshift consistently
with passive evolution of the stellar populations in Es: however, in
our discussion about the position of the end--products with respect to
the FP and the FJ, for simplicity the mass--to--light ratio $\mlk
\equiv \Mstar/\Lk$ is maintained constant in the progenitors and in
the end--products. Thus, in order to place the isotropic
one--component progenitors ($\Re\simeq4$ kpc, $\Mstar=4 \times 10^{11}
\Msol$, $\sg0\simeq259~{\rm km\,s^{-1}}$) on the FP, we assume
$\mlk\simeq 1.5$ (in the K-band), while the anisotropic one--component
($\Re\simeq4$ kpc, $\Mstar=4 \times 10^{11} \Msol$, $\sg0\simeq
271~{\rm km\, s^{-1}}$) and the two--component ($\Re\simeq4$ kpc,
$\Mstar=2 \times 10^{11} \Msol$, $\sg0\simeq279~{\rm km\, s^{-1}}$)
models require $\mlk\simeq 1.4$ and $\mlk\simeq 0.65$, respectively.
With this choice, the progenitors are placed in all simulations on the
filled circle at the bottom left of Fig.~4, where equation (10) with
its scatter is shown.

The position of the end--products (generally not spherically
symmetric) in the parameter space where the FP is defined depends on
the line--of--sight direction: as a consequence, each end--product,
owing to projection effects, determines a two dimensional region in
Fig.~4 where it is represented by a set of points corresponding to 8
random projections. A first interesting result is that {\it the
projection effects, though important, are not larger than the observed
FP scatter}, in accordance with other numerical and analytical
explorations (NLC03a, Lanzoni \& Ciotti 2003).  In addition, the
behavior of the end--products with respect to the FP shows a certain
dependence on the characteristics of the initial galaxies. In
particular, although {\it the accordance with the FP is remarkable for
all our simulations} (in agreement with what found in case of binary
merging; NLC03a, Nipoti, Londrillo \& Ciotti 2003b, Dantas et
al. 2003, Gonzalez-Garcia \& van Albada 2003) the end--products of
isotropic one--component and two--component galaxy models (empty
circles and crosses in Fig.~4, respectively) stay preferentially below
the FP best--fit, while most of the end--products of anisotropic
one--component galaxies (empty squares) are found above the FP
best--fit.

As well known, the fact that a galaxy lies on the edge--on FP does not
imply that it satisfies the FJ relation too, because the latter
contains also information about the position of Es on the face--on
FP. In Fig.~5 we plot the position of the end--products with respect
to the near infra-red FJ relation (solid line): the central velocity
dispersion and the luminosity are normalized to those of the
progenitors, which are placed at the origin, while the vertical bars
in the diagram indicate the range in $\sg0$ associated to each
end--product, owing to projection effects. Figure~5 shows that, as for
the FP, the end--products of the simulations {\it do reproduce well
the observed FJ}. This behavior could be considered at variance with
what found in NLC03a and Nipoti et al. (2003b), where it was shown
that in merging hierarchies the edge--on FP is usually well reproduced
but the FJ is not, in the sense that the end--products are
characterized by too low $\sg0$ for their mass (luminosity). However,
it should be noted that the (expected!) deviation from the FJ as a
consequence of dissipationless merging becomes apparent only after
several steps of merging. Indeed, NLC03a found the merging
end-products to be in accordance with the FJ relation when restricting
to only one or two merging events. 

\begin{figure}
\begin{center}
\parbox{1cm}{ \psfig{file=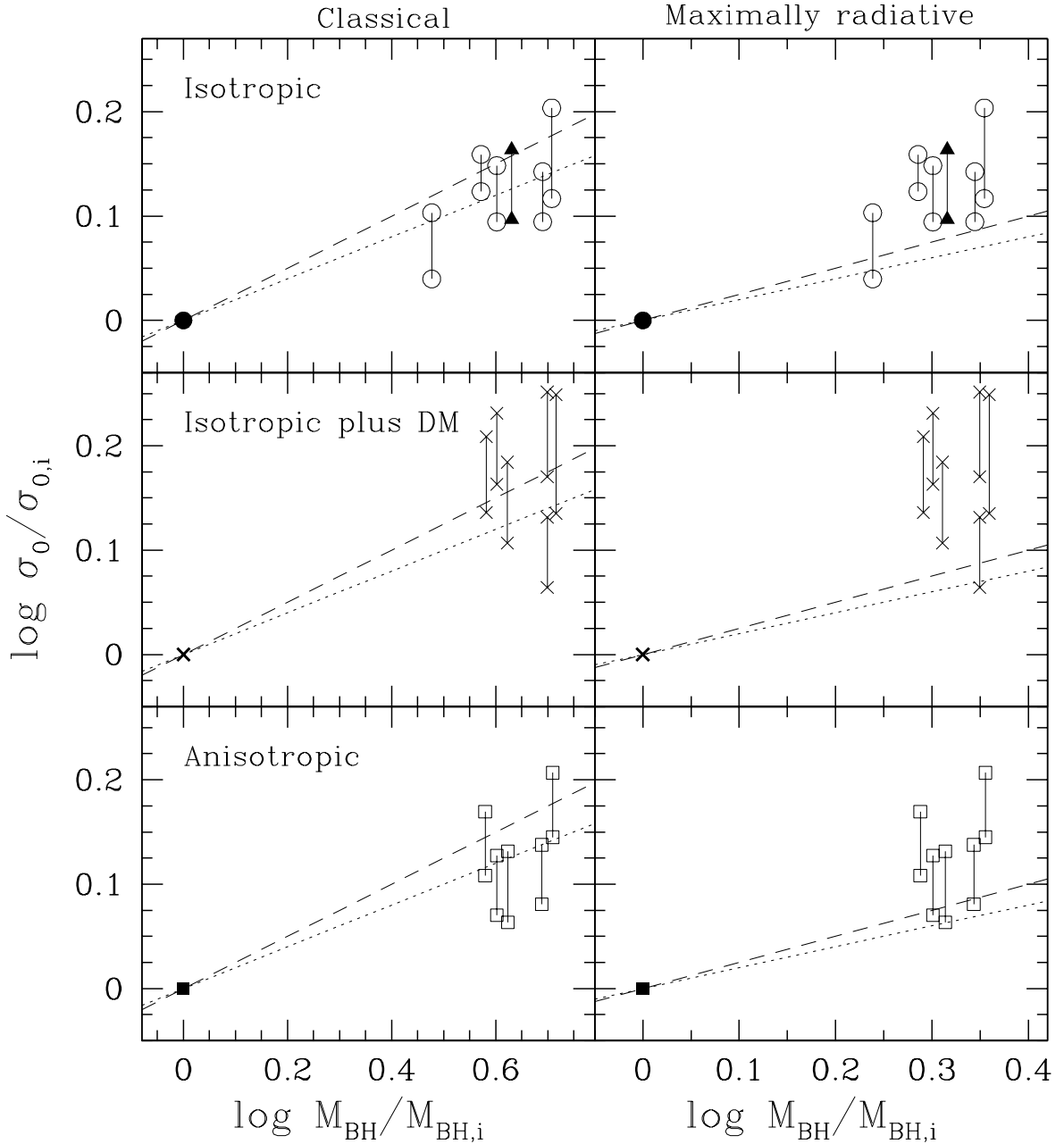,height=17cm}}
\caption{ {\it Left panels}: galactic central velocity dispersion vs.
BH mass for classical BH merging, in case of isotropic progenitors
(upper panel), progenitors with DM halo (intermediate panel), and
anisotropic progenitors (lower panel); $\sigma_{\rm 0,i}$ and $M_{\rm
BH,i}$ are the central velocity dispersion and BH mass of the
progenitors, respectively. The simulations shown and the symbols are
the same as in Figs 4 and 5. The bars indicate the range spanned by
projection effects: as in Fig.~5, each group of bars is artificially
split. Dashed and dotted lines represent the $\Mbh$-$\sg0$ relation
(equation~12) for $\alpha=4$ and $\alpha=5$, respectively.  {\it Right
panels}: same data as in left panels, but for maximally radiative BH
merging.}
\label{appenfig}
\end{center}
\end{figure}

\subsection{Multiple merging and the $\Mbh$-$\sg0$ relation}

As outlined in the Introduction, a question naturally raised in the
considered scenario is whether the $\Mbh$-$\sg0$ relation is preserved
by multiple dissipationless merging. This relation, between  the
mass of the central BH and the central velocity dispersion of the
host galaxy or bulge, can be written in the form
\begin{equation}
\Mbh \propto \sg0^{\alpha},
\end{equation}
where the exact value of $\alpha$ is still matter of debate, but seems
to be in the range $4 - 5$ (see, e.g., Gebhardt et al.  2000;
Ferrarese \& Merritt 2000, Merritt \& Ferrarese 2001, Tremaine et
al. 2002). As in NLC03a, here we try to get some indications about the
effects of multiple mergings on the $\Mbh$-$\sg0$ relation, by using
the results of our numerical simulations, though in them we do not
explicitly take into account the presence of BHs. The plausibility of
our approach is due to the fact that at the equilibrium the presence
of a BH has no significant influence on $\sg0$, as the sphere of
influence of a central BH with mass $\Mbh$ has a fiducial radius $\rbh
\equiv G \Mbh / \sg0^2$, about one order of magnitude smaller than the
typical aperture radius used to determine $\sg0$. In addition,
Milosavljevic \& Merritt (2001) showed that the BH binary, a natural
consequence of galaxy merging, though modifying the inner density
profile, does not affect significantly the projected central velocity
dispersion measured within standard apertures.

On the basis of these considerations, we simply assume that each of
the five galaxies contains a BH, whose mass is related to the galaxy
central velocity dispersion by equation (12), and the merger remnant
contains a BH obtained by the merging of the BHs of the
progenitors. As in Ciotti \& van Albada (2001) and NLC03a, we consider
two extreme situations for the BH mass addition: the case of {\it
classical} combination of masses ($M_{\rm BH,1+2}=M_{\rm BH,1}+M_{\rm
BH,2}$, with no emission of gravitational waves), and the case of {\it
maximally efficient radiative merging} of two non--rotating BHs
($M^2_{\rm BH,1+2}=M^2_{\rm BH,1}+M^2_{\rm BH,2}$). Following this
choice, in Fig.~6 we plot the central velocity dispersion of the
mergers versus the mass of the resulting BH, in the case of classical
(left panels) and maximally radiative (right panels) BH merging. In
the diagrams the dashed and dotted lines correspond to $\alpha=4$ and
$\alpha=5$ in equation (12), respectively. We note that, owing to the
(relatively) small range of galaxy masses explored, the difference
between the values of $\sg0$ predicted for these two values of the
exponent are always smaller than the projection effects on $\sg0$ of
the models (vertical bars): for this reason, our considerations will
be in practice independent of the exact value of $\alpha$. 

We start up considering the case of classical combination of BH masses
(Fig.~6, left panels).  In this case it is obvious that, due to the
striking similarity between the exponents of the $\Mbh$-$\sg0$ and the
FJ relations, all the comments made in Section 5.2.3 apply also here.
The situation changes substantially if maximally radiative BH merging
is considered (Fig.~6, right panels): in this case the BH mass does
not increase linearly with the (stellar) galaxy mass and, as a
consequence, the $\Mbh$-$\sg0$ relation would predict a lower $\sg0$
for the merger remnant of given luminosity, with respect to the
classical addition case. It is important to note that in the present
exploration the behavior of the mergers is in better accordance with
the $\Mbh$-$\sg0$ relation in case of a classical addition law than in
the case of a maximally radiative BH merging, again at variance with
the results presented in NLC03a. The origin of this seemingly
different behavior can be again traced back directly to the
preservation of the FJ relation discussed in Section 5.2.3.

All the results presented in this Section are based on the assumption
that the BHs of all the merging galaxies will contribute to the
formation of the final BH, leading of course to an oversimplified
scenario. In fact, there are at least two basic mechanisms that could
be effective in expelling the central BHs in a binary or multiple
merging. The first is the slingshot effect: if a third galaxy is
accreted by a merger remnant still hosting a BH binary, then the
escape of one of the three BHs is highly possible (see, e.g., Haehnelt
\& Kauffmann 2002, Volonteri, Haardt \& Madau 2003). Clearly, this
process could be particularly effective in the considered situation of
multiple merging: from our simulations it results that multiple
mergings are expected to happen in few Gyrs, i.e., with time-scales
hardly longer than those of BH merging (see, e.g., Yu 2002). A second
physical mechanism that could produce the ejection of the resulting BH
is the so--called ``kick-velocity'' effect: if, in a gravitationally
radiative BH merging, a fraction (even a few thousandths) of the mass
of the BH binary is emitted {\it anisotropically} as gravitational
waves, the recoil due to linear momentum conservation is sufficient to
expel the two merging BHs from the remnant (see, e.g., Flanagan \&
Hughes 1998, Ciotti \& van Albada 2001). We note that, in any case, a
substantial amount of BH ejection during merging would necessarily
lead to the violation of the observed linear relation between the mass
of the central BH and the mass of the host bulge or galaxy (the
so--called Magorrian relation; Magorrian et al. 1998). Thus, the
preservation of the $\Mbh$-$\sg0$ and of the Magorrian relations
represents an important (and difficult) astrophysical problem related
to the formation of BCGs.

\subsection{Metallicity gradients}

\begin{figure}
\begin{center}
\parbox{1cm}{ \psfig{file=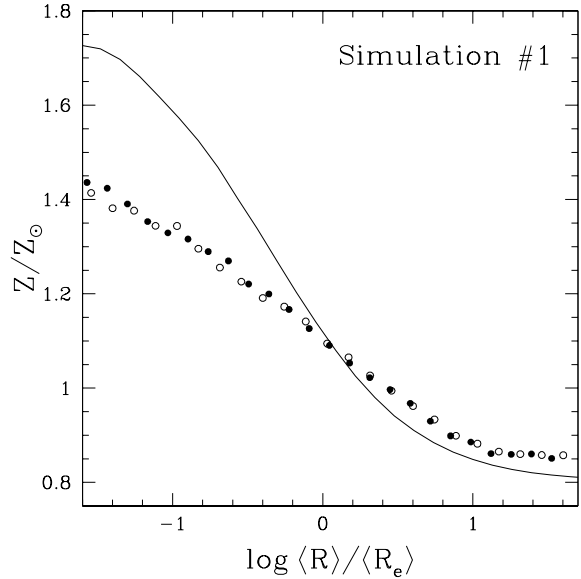,width=8cm}}
\caption{Projected metallicity versus circularized radius. Solid line:
initial metallicity distribution of each progenitor galaxy. Filled and
empty symbols: metallicity distribution of the end--product of
simulation~\#1 projected along the major and minor axis,
respectively.}
\label{appenfig}
\end{center}
\end{figure}

As it is well known, metallicity gradients are a common feature of Es
(Peletier 1989, Carollo et al. 1993), and their robustness in the
context of galaxy merging was early recognized as an important
constraint for scenarios of galaxy formation. For example, White
(1980) found that the remnant of the merging of two equal mass
galaxies has a metallicity gradient $\sim$ 20 per cent smaller than
its progenitors. This result suggested that in a scenario of
(dissipationless) hierarchical merging the gradients could be erased
in few subsequent mergings.  It is interesting to extend these
considerations to the case of multiple merging like that analysed in
this work, even if in the present case their constraining power is
substantially reduced. In fact, BCGs normally reside at the center of
clusters, where metal rich intracluster medium flows (such as cooling
flows) alter considerably the metallicity distribution.

From a dynamical point of view, the observed projected gradients
correspond to phase--space projections of the intrinsic metallicity
distribution of stars in their orbits. Ciotti, Stiavelli \& Braccesi
(1995) described a simple technique to derive the intrinsic
metallicity distribution in the case of spherically symmetric
galaxies.  By adopting the same approach, in the initial conditions of
the one--component simulations we assigned to each stellar particle a
metallicity given by
\begin{equation}
{Z \over \Zsun}=1.3 {Q \over \Psi(0)}+0.8,
\end{equation}
where $\Zsun$ is the solar metallicity, $\Psi(0)$ is the central
relative potential of each progenitor, and $Q$ is defined below
equation (5). This choice corresponds to a projected central
metallicity $Z(R=0)\simeq2\Zsun$. In our investigation we assume that
the metallicity of each particle remains constant during the dynamical
evolution of the system.  We quantified the projected metallicity
gradients, in both the initial galaxies and the end--products, by
measuring the ``center-to-edge metallicity difference'' as defined by
White (1980):
\begin{equation}
\Delta Z={ {\langle Z \rangle _{\rm in}} - {\langle Z \rangle _{\rm out}} \over {\langle Z \rangle }},
\end{equation}
where ${\langle Z \rangle}$, ${\langle Z \rangle _{\rm in}}$, and
${\langle Z \rangle _{\rm out}}$ are the metallicities averaged over
the whole distribution, inside the projected radius enclosing 1/3 of
the total mass, and outside the projected radius enclosing 2/3 of the
total mass, respectively.  We measured $\Delta Z$ of the merger
remnants of isotropic and anisotropic one--component progenitors,
considering the projections along the three principal axes. In all the
cases we found that it is significantly lower in the end--products
than in the progenitors: the decrease in $\Delta Z$ is in the range
$30 - 60\%$ of the initial value.  This result is presented with an
example in Fig.~7, where we plot the projected metallicity as a
function of the circularized radius for the end--product of the
merging of 4 galaxies (simulation~\#1), and, for comparison, for its
progenitors. On the basis of the described results, we can conclude
that in case of multiple merging the observed reduction of the
metallicity gradient is consistent with the repeated application of
``White's 20\% rule''. Note also that, if the merging galaxies contain
a substantial gas fraction, the metallicity gradient could be restored
by a significant star formation event. Interestingly, Carollo et
al. (1993) find that the correlation between mass and the metallicity
gradient fails for high mass Es, indicating that in some cases very
massive galaxies have smaller gradient than less luminous Es.

\section{Discussion and conclusions}

From a theoretical point of view, it is well known that galactic
cannibalism can be an effective mechanism for the formation of BCGs in
general and cDs in particular. On the other hand, just circumstantial
evidence of this process comes from the observations, and only of the
phase after merging (for example, BCGs with multiple nuclei). In this
paper we explored, with the aid of N-body simulations based on the
observationally available phase--space information, the hypothesis
that the group of five Es in the core of the X-ray galaxy cluster
C0337-2252 is a strong candidate for the galactic cannibalism
scenario. We summarize below the main results and then we discuss,
with the aid of a few ``ad hoc'' simulations, a couple of important
points raised by the presented results.

\begin{itemize}

\item In all of the explored cases at least 3 galaxies merge before
$z=0$. For some values of the parameters all the 5 galaxies are
involved in the merging. The number of merging galaxies depends on the
cluster structure and, in some cases, also on the particular
realization of the initial condition for a given cluster model.  It is
shown that the driving mechanism of the merging process is the
dynamical friction of the galaxies against the diffuse cluster DM: if
the live halo is substituted by a fixed potential the number of
merging is drastically reduced.

\item The merger remnants are always similar in their main structural
and dynamical properties to a real BCG. Their SB profiles are well
represented by the de Vaucouleurs law up to $\sim 10\cRe$, with no
evidence of the diffuse and extended halo typical of cD galaxies (see
discussion below).

\item It is found that the merging end--products nicely follow the
K-band FP and FJ relations, under the hypothesis that the five
galaxies are placed on these two scaling relations, by an appropriate
choice of the stellar mass-to light ratio. These results are only
weakly dependent on the specific structure and dynamics of the galaxy
models used as initial conditions (i.e., presence of DM, orbital
anisotropy).

\item The behavior of the end--products with respect to the
$\Mbh$-$\sg0$ relation depends on the details of the BH merging.
Assuming that each galaxy initially hosts at its center a supermassive
BH (whose mass follows the observed $\Mbh$-$\sg0$ relation) and that
all the involved BHs finally merge, we found that the $\Mbh$-$\sg0$
relation is preserved if the BH masses add classically (in accordance
with the results on the FJ relation), while the end--products lie
systematically above the observed $\Mbh$-$\sg0$ relation in case of
substantial emission of gravitational waves.

\item The metallicity gradient in the remnants is $30 - 60 \%$ lower
than in the initial galaxies. This result is consistent with the
results of binary merging obtained by White (1980), and also with
observations (Carollo et al. 1993), reporting a range of metallicity
gradients in the most massive Es.

\end{itemize}

Thus, the results presented in this paper strongly suggest that {\it
the observed system of five galaxies in the galaxy cluster C0337-2252
could be a BCG in formation in the phase before merging}. However, as
already pointed out, this quite successful scenario is unable to
reproduce the diffuse low luminosity halo of cD galaxies, which are a
substantial fraction of BCGs. It has been speculated that the
interaction of a central massive galaxies (like that produced in the
considered multiple merging) with smaller and lower density galaxies
could be responsible of its formation; another dynamical aspect that
could be important in establishing the cD profile is the cosmological
collapse of the DM cluster distribution during the galactic
cannibalism event. In order to qualitatively address these two
important points, we also ran 5 additional simulations, denominated as
\#3c,cc, \#17c,cc, and \#1s in Table~2. In particular, in simulation
\#1s we used the same initial condition as in \#1, but we also added a
population of 50 smaller galaxies (again modeled as one--component
Hernquist models with $\Mstar=5 \times 10^{10}\Msol$, $\rcstar \simeq
0.55$ kpc, and $\Nstar=256$). This additional population was
distributed in the cluster by extraction from the cluster DF. Two are
the main results of simulation \#1s: the first is the fact that the
number of merging galaxies is still 4, but also 29 smaller galaxies
are cannibalized at the cluster center. Thus, the total mass of the
end--product is approximately a factor of 1.9 larger than the
corresponding end--product of simulation \#1. The second result is the
fact that the SB profile in \#1s does not present the CD extended
halo, while the end--product still nicely obeys the FP and FJ
relations (diamonds in Figs 4, 5).  In the other 4 simulations we
explore a few cases in which at $z=0.59$ the cluster is collapsing. In
particular, we reran simulations \#3 and \#17, assuming an initial
virial ratio for the cluster DM component $2T/|W|=0.8$ (\#3c, \#17c)
and $2T/|W|=0.5$ (\#3cc, \#17cc).  Interestingly, in simulations
\#3cc, \#17c, \#17cc, the number of merging galaxies increases with
respect to the case of a virialized cluster (see Table~2).  In any
case, we found that the structural and dynamical properties of these
four merger remnants, as well as their behavior with respect to the
FP, the FJ and the $\Mbh$-$\sg0$ relations (triangles in Figs 4, 5,
6), are similar to that of the other isotropic galaxy models
explored. Again, while the general picture of the paper is confirmed,
the simulations fail to reproduce the characteristic cD envelope.

We could then ask whether {\it dissipationless} merging, which works
quite well in the presented case of BCG formation, could be a
universal mechanism for the formation of Es in general. NLC03a showed
that it is unlikely that (binary) {\it dissipationless} galaxy merging
is the dominant mechanism for the formation of normal Es, as it fails
to reproduce some of their observed scaling relations over a large
range in luminosity. However, NLC03a numerical simulations, as well as
ours (indicating that many properties of the BCGs are reproduced by
multiple merging of {\it a few} luminous Es), show that {\it few}
merging events are compatible with the existence of the observed {\it
thin} FP of Es, though the remnants have in general lower central
velocity dispersion and larger effective radius, with respect to real
Es. Remarkably, real BCGs do follow quite closely the FP, and, on the
other hand, many of them have lower $\sg0$ and mean effective SB than
predicted by the FJ and the radius-luminosity relation of normal Es,
respectively (see, e.g., Oegerle \& Hoessel 1991).

Thus, we conclude that, while all the simulations we ran strongly
support {\it one} of the two ingredients necessary for the formation
of a cD galaxy, namely the galactic cannibalism at the galaxy center,
they are unable to reproduce the peculiar extended halo of cDs. This
is not at variance with previous works that showed that the halo
formation is a more ``delicate'' dynamical phenomenon than the
straight galaxy merging. It should be also recalled that observations
suggest that cDs seem to be common only at the center of rich
clusters, while in small clusters BCGs have a $R^{1/4}$ luminosity
profile over a large radial range (Thuan \& Romanishin 1981). Perhaps
very high resolution simulations of satellite accretion in different
environments will be able to reveal the dynamical conditions necessary
for the formation of cDs, along a line of research already started
(see, e.g., Athanassoula et al. 2001).

\section*{Acknowledgments}

We would like to thank Andrew~Benson, Oleg~Gnedin, Pasquale~Londrillo,
Jeremiah~Ostriker, and the anonymous referee for their useful comments
on the manuscript. C.N. is grateful to STScI (Baltimore) for its
hospitality, and to CINECA (Bologna) for assistance with the use of
the Cray T3E and of the IBM Linux Cluster. C.N. was supported by the
STScI Collaborative Visitors Program; L.C. by MURST Cofin
2000. M.S. is partially supported by NASA NAG5-12458.

\end{document}